\documentclass[twocolumn]{revtex4}

\usepackage{graphicx}
\usepackage{dcolumn}
\usepackage{amsmath}

\begin{document}

\title{Incommensurate Antiferromagnetism Coexisting with Superconductivity
in Two-Dimensional d-p Model
}

\author{Takashi Yanagisawa$^{a,c}$, Mitake Miyazaki$^b$, and Kunihiko Yamaji$^{a,c}$} 

\affiliation{$^a$Condensed-Matter Physics Group, Nanoelectronics Research 
Institute, National Institute of Advanced Industrial Science and Technology (AIST),
Tsukuba Central 2, 1-1-1 Umezono, Tsukuba 305-8568, Japan\\
$^b$Hakodate National College of Technology, 14-1 Tokura, Hakodate, Hokkaido 042-8501, 
Japan\\
$^c$CREST, Japan Science and Technology Agency (JST), Kawaguchi, 
Saitama 332-0012, Japan 
}


\begin{abstract}
Numerical studies of the two-dimensional d-p model using the Gutzwiller ansatz  
have exhibited
the incommensurate antiferromagnetic state coexisting with superconductivity
in the under- and lightly doped regions.
Our results are based on the variational Monte Carlo method for the three-band
Hubbard model with d and p orbitals.
We obtained the finite superconducting condensation energy for the 
coexistent sate at the doping rate $x=1/8$, 1/12, and 1/16, up to the systems of 
256 unit cells with 768 atoms (oxygen
and copper atoms).
The phase diagram for the hole-doped case is consistent
with recent results reported for layered high temperature cuprates.
\end{abstract}

\maketitle

The mechanisms of 
superconductivity (SC) in high-temperature superconductors
have been extensively studied using various two-dimensional 
(2D) models of electronic interactions\cite{dag94,ben03,and97,mor00}.  
It is of primary importance to clarify the phase diagram, particularly 
the electronic state in the underdoped region
adjacent to the antiferromagnetic (AF) phase, termed the pseudo-gap phase.
It is unclear whether the phase diagram for La$_{2-x}$Sr$_x$CuO$_4$ is
intrinsic for high-$T_c$ cuprates or not, although it is
often recognized as a typical phase diagram.  It is sometimes declared that
disorder effects play some role in the spin glass phase of
La$_{2-x}$Sr$_x$CuO$_4$.
Thus, it is fair to say that the phase diagram has never been clarified.

The 2D three-band d-p model is the most fundamental model for high-temperature
cuprates\cite{hir89,sca91,tak97,gue98,kob98,koi00,yan01}.
Although we have a solution of the gap equation within a weak coupling perturbation
theory in the limit $U\rightarrow 0$\cite{koi01,yan08}, 
it is, however, extremely hard to show the possibility of superconductivity
exactly for finite and large Coulomb repulsion.  
Thus we adopt the Gutzwiller ansatz for
the wave function and examine the ground state within the space of
variational functions.
We employ the variational Monte Carlo method\cite{gro87,yok87,nak97,yam98} 
to evaluate the expectation 
values of several physical properties.

The purpose of this study is to investigate the coexistence of superconductivity
and antiferromagnetism for the 2D d-p model.
We have found that the coexistent state has indeed the
lowest energy in the variational space at the doping rate $x=0.125$, 0.08333,
and 0.0625 in the low-doping region.
At $x=0.125$, the incommensurate antiferromagnetic
state has eight-lattice periodicity, as reported on the basis of neutron scattering
measurements\cite{tra96}. 
The periodicity increases as $x$ decreases; we have twelve lattice periodicity
at $x=0.0833$ and sixteen-lattice periodicity at $x=0.0625$.


The Hamiltonian is the d-p model containing the on-site Coulomb 
repulsion for d electrons and is written as\cite{eme87}

\begin{eqnarray}
H_{dp}&=& \epsilon_d\sum_{i\sigma}d_{i\sigma}^{\dag}d_{i\sigma}
+ \epsilon_p\sum_{i\sigma}(p_{i+\hat{x}/2\sigma}^{\dag}p_{i+\hat{x}/2\sigma}
\nonumber\\
&+& p_{i+\hat{y}/2\sigma}^{\dag}p_{i+\hat{y}/2\sigma})
\nonumber\\
&+& t_{dp}\sum_{i\sigma}[d_{i\sigma}^{\dag}(p_{i+\hat{x}/2\sigma}
+p_{i+\hat{y}/2\sigma}-p_{i-\hat{x}/2\sigma}-p_{i-\hat{y}/2\sigma})\nonumber\\
&+& {\rm h.c.}]+t_{pp}\sum_{i\sigma}[p_{i+\hat{y}/2\sigma}^{\dag}p_{i+\hat{x}/2\sigma}
-p_{i+\hat{y}/2\sigma}^{\dag}p_{i-\hat{x}/2\sigma}\nonumber\\
&-&p_{i-\hat{y}/2\sigma}^{\dag}p_{i+\hat{x}/2\sigma}
+p_{i-\hat{y}/2\sigma}^{\dag}p_{i-\hat{x}/2\sigma}+{\rm h.c.}]\nonumber\\
&+& U_d\sum_i d_{i\uparrow}^{\dag}d_{i\uparrow}d_{i\downarrow}^{\dag}
d_{i\downarrow}.
\end{eqnarray}
$d_{i\sigma}$ and $d^{\dag}_{i\sigma}$ are the operators for the $d$ electrons.
$p_{i\pm\hat{x}/2\sigma}$ and $p^{\dag}_{i\pm\hat{x}/2\sigma}$ denote the 
operators for the $p$ electrons at the site $R_{i\pm\hat{x}/2}$, and in a
similar way, $p_{i\pm\hat{y}/2\sigma}$ and $p^{\dag}_{i\pm\hat{y}/2\sigma}$  
are defined.  $U_d$ is the strength of the on-site Coulomb energy between
$d$ electrons.  
The number of sites is denoted as $N_s$, and the total number of atoms is
$N_a=3N_s$.
The total number of fermions is denoted as $N_e$.
The energy unit is given by $t_{dp}$ in this paper.

\begin{figure}
\includegraphics[width=\columnwidth]{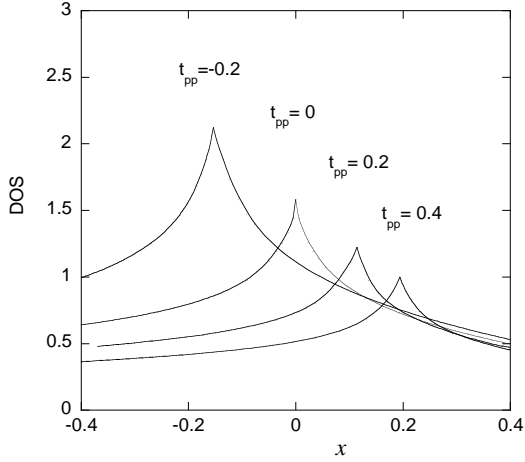}
\caption{
Density of states of the d-p model as a function of the carrier density
$x$ for $t_{pp}=0.4$, 0.2, 0.0 and -0.2.
We set $\epsilon_p=0$, $\epsilon_d=-1$ and $t_{dp}=1$ (energy unit).
}
\label{dos-dp}
\end{figure}

The van Hove singularity in the density of states plays an important role 
in two-dimensional models.
We define the density of states as
\begin{equation}
D(\epsilon)= \frac{1}{N_s}\sum_{{\bf k}}\delta(\epsilon-\xi_{{\bf k}}),
\end{equation}
where $\xi_{{\bf k}}=\epsilon_{{\bf k}}-\mu$ ($\mu$ is the Fermi energy)
and $\epsilon_{{\bf k}}$ is the band crossing the Fermi energy.
We examine the hole-doped case within the hole picture where the lowest band 
is occupied up to the Fermi energy $\mu$.
For this purpose, we employ the electron-hole transformation
$t_{pp}\rightarrow -t_{pp}$, $t_{dp}\rightarrow -t_{dp}$, 
and we set $\epsilon_p-\epsilon_d>0$.
The density of states
$D(\epsilon)$ as a function of the carrier density $x$ is shown in 
Fig. \ref{dos-dp} for $t_{pp}=0.4$, 0.2, 0, and -0.2.
$x=0$ corresponds to the half-filled band. 
For $t_{pp}=0$, the van Hove singularity is at $x=0$.
It moves to the hole-doped side of $x>0$ for $t_{pp}>0$ and to the 
electron-doped side for $t_{pp}<0$.
We have the van Hove singularity at $x\sim 0.16$ for $\epsilon_p-\epsilon_d=2$
and $t_{pp}=0.4$.
Thus we set parameters to be $\epsilon_p-\epsilon_d=2$ and $t_{pp}=0.4$
in  the main computations, and $U_d=8$ in this paper. 
This is in good accordance with the results of cluster 
estimations\cite{esk89,hyb90,mcm90}.
The van Hove singularity approaches $x=0$ as the level difference
$\epsilon_p-\epsilon_d$ becomes large.
Hence, we expect that the critical temperature $T_c$ has a peak as a function of 
$\epsilon_p-\epsilon_d$ if we fix the carrier density $x$.


We adopt the Gutzwiller ansatz for the ground-state wave function $\psi$:
$\psi=P_G\psi_0$, 
where $\psi_0$ is a trial one-body wave function and
\begin{equation}
P_G=\prod_i(1-(1-g)n_{di\uparrow}n_{di\downarrow})
\end{equation}
is the Gutzwiller projection operator.
$g$ is the variational parameter in the range of $0\leq g\leq 1$.
The wave function considered in this paper is a coexistent state which
is given by the solution of the Bogoliubov-de Gennes equation:
\begin{equation}
\sum_j (H_{ij\uparrow}u_j^{\lambda}+F_{ij}v_j^{\lambda})=E^{\lambda}u_i^{\lambda},
\end{equation}
\begin{equation}
\sum_j (F_{ji}^*u_j^{\lambda}-H_{ji\downarrow}v_j^{\lambda})=E^{\lambda}v_i^{\lambda},
\end{equation}
for a trial Hamiltonian $H_{ij\sigma}$ and $F_{ij}$,
where $(H_{ij\sigma})$ and $(F_{ij})$ are $3N_s\times 3N_s$ matrices including 
the terms for $d$, $p_x$, and $p_y$ orbitals.
The Bogoliubov operators are written as
\begin{equation}
\alpha_{\lambda}=\sum_i(u_i^{\lambda}a_{i\uparrow}
+v_i^{\lambda}a_{i\downarrow}^{\dag})~~ (E^{\lambda}>0),
\end{equation}
\begin{equation}
\alpha_{\bar{\lambda}}=\sum_i(u_i^{\bar{\lambda}}a_{i\uparrow}
+v_i^{\bar{\lambda}}a_{i\downarrow}^{\dag})~~ (E^{\bar{\lambda}}<0).
\end{equation}
$a_{i\sigma}$ denotes $d_{i\sigma}$, $p_{i+\hat{x}/2\sigma}$, and
$p_{i+\hat{y}/2\sigma}$ corresponding to the components of $u_i^{\lambda}$
and $v_i^{\lambda}$.
The coexistent superconducting state is\cite{him02,miy02}
\begin{eqnarray}
\psi&=& P_N\prod_{\lambda}\alpha_{\lambda}\alpha_{\bar{\lambda}}^{\dag}
|0\rangle\nonumber\\
&=& {\rm const.}P_N{\rm exp}\left(-\sum_{ij}\phi_{ij}a_{i\uparrow}^{\dag}
a_{j\downarrow}^{\dag}\right)|0\rangle,
\end{eqnarray}
where $|0\rangle$ is the vacuum state annihilated by $d_{i\sigma}$, 
$p_{i+\hat{x}/2\sigma}$, and $p_{i+\hat{y}/2\sigma}$.
Since $\psi_{SC}$ satisfies $\alpha_{\lambda}\psi_{SC}=0$, using the
Hausdorff formula, $\phi_{ij}$ is
determined as
\begin{equation}
\phi_{ij}= (U^{-1}V)_{ij},
\end{equation}
where we define the matrices $U$ and $V$ as $U_{\lambda j}=u_j^{\lambda}$ and
$V_{\lambda j}=v_j^{\lambda}$.
$P_N$ fixes the electron number to be $N_e$.
The antiferromagnetic order parameter is contained in $(H_{ij\sigma})$ and the 
superconducting gap function is in $(F_{ij})$.

Since the incommensurate state was shown to be stable in the lightly doped
region, we assume the spatial variation for the order parameters.
The trial Hamiltonian is the Hartree-Fock Hamiltonian given 
as\cite{gia90,yan02,yan03,miy04}
\begin{equation}
H_{trial}= K+\sum_{i\sigma}[\delta n_{di}-\sigma(-1)^{x_i+y_i}m_i]
d_{i\sigma}^{\dag}d_{i\sigma}.
\label{htri}
\end{equation}
Corresponding to the energy levels $\epsilon_d$ and $\epsilon_p$,  variational
parameters $\tilde{\epsilon_p}$ and $\tilde{\epsilon_d}$ are incorporated in the
noninteracting part K in eq. (\ref{htri}).
We assume the spatial variations to be
\begin{equation}
\delta n_{di}=-\sum_j \frac{\alpha}{{\rm cosh}(x_i-x_j^{inc})},
\end{equation}
\begin{equation}
m_i= \Delta_{inc}\prod_j{\rm tanh}(x_i-x_j^{inc}),
\end{equation}
for parameters $\alpha$, $\Delta_{inc}$, and $x_j^{inc}$.
$x_{j}^{inc}$ determines the periodicity of oscillation; we set
$x_j^{inc}=j/(2x_v)$ for the variational parameter $x_v$.
The energy is computed for several values of $x_v$ such $x_v=1/4$, 1/8, $\cdots$.
A small spatial charge oscillation, which is, at most, ten percent of the
total density, is induced owing to the oscillation potential $\delta n_{di}$ and
$m_i$\cite{miy04}.  Thus we assume the following superconducting order parameter:
\begin{equation}
\Delta_{i,i+\hat{x}}= \Delta_x {\rm cos}[Q_{\delta}(x_i+\hat{x}/2)],
\end{equation}
\begin{equation}
\Delta_{i,i+\hat{y}}= \Delta_y {\rm cos}(Q_{\delta}x_i),
\end{equation}
for $Q_{\delta}=2\pi x_v$.
We assume the d-wave symmetry for the SC gap function:
$\Delta_x=-\Delta_y\equiv \Delta$.
The superconducting order parameter oscillates so that the amplitude has
a maximum in the hole-rich region and a minimum in hole-poor region.
The energy expectation value
$E=\langle\psi |H|\psi\rangle/\langle\psi |\psi\rangle$
is evaluated using a Monte Carlo Metropolis
algorithm, which is a standard method in variational Monte Carlo
computations.

\begin{figure}
\includegraphics[width=\columnwidth]{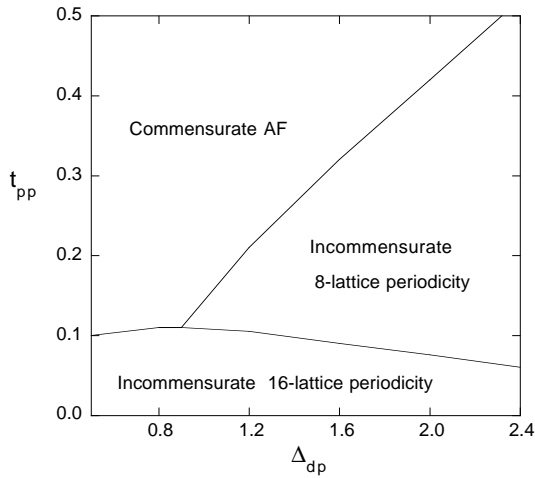}
\caption{
Phase diagram of stable antiferromagnetic state in the plane of
$\Delta_{dp}=\epsilon_p-\epsilon_d$ and $t_{pp}$ obtained for
$16\times 4$ lattice.
}
\label{phase-inc}
\end{figure}

\begin{figure}
\includegraphics[width=\columnwidth]{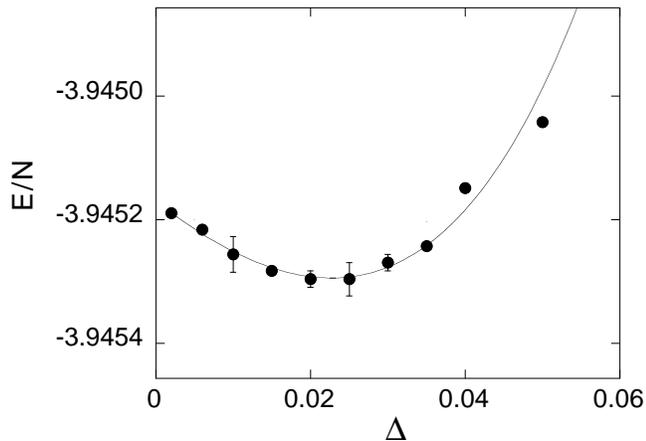}
\caption{
Energy of the coexistent state as a function of the SC order parameter
for $x=0.125$ on $16\times 4$ lattice.
We assume the incommensurate antiferromagnetic order (stripe).
Parameters are $\epsilon_p=0$, $\epsilon_d=-2$ and $t_{pp}=0.4$.
}
\label{E-16x4}
\end{figure}

\begin{figure}
\includegraphics[width=\columnwidth]{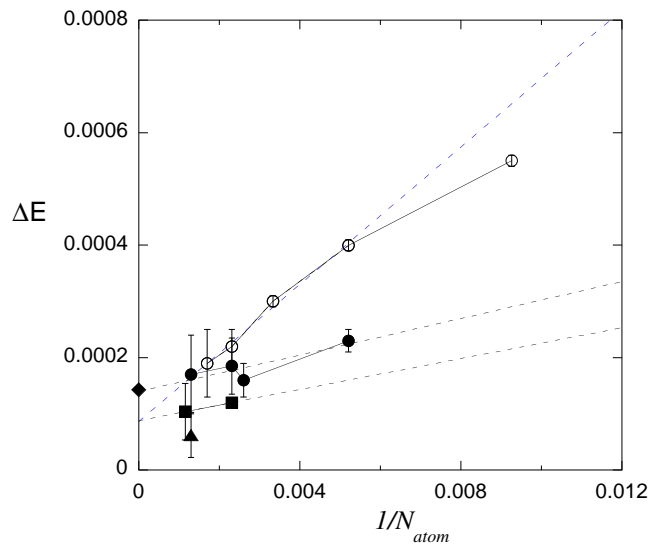}
\caption{
Energy gain due to the SC order parameter as a function of the system
size $N_{atom}=3N_s$.
Parameters are $\epsilon_p=0$, $\epsilon_d=-2$, $t_{pp}=0.4$, and $U_d=8$.
The open circles are for the simple $d$-wave pairing at the hole density $x=0.2$.
The solid symbols indicate the energy gain of the coexistent state:
the solid circles are those at $x=0.125$, solid
squares are those at $x=0.08333$ and the solid triangle is that at $x=0.0625$.
The diamond shows the SC condensation energy obtained on the basis of
specific heat measurements on the optimally doped 
YBa$_2$Cu$_3$O$_{6+x}$ at $x=0.92$\cite{lor93}.
}
\label{dE-N}
\end{figure}

\begin{figure}
\includegraphics[width=\columnwidth]{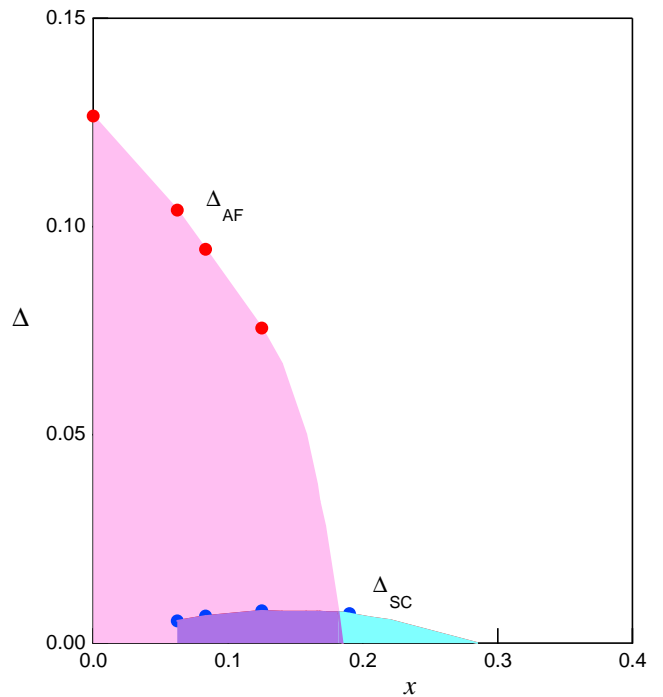}
\caption{
Phase diagram of the d-p model based on the Gutzwiller wave function.
}
\label{d-p-phase}
\end{figure}


The condensation energy $E_{cond}$ is defined
as the difference  $E_{cond}=E(\Delta\rightarrow 0)-E(\Delta)$ for the optimized
energy.
The energy of the antiferromagnetic state would be lowered further if we consider
the incommensurate spin correlation in the wave function.
The phase diagram in Fig. \ref{phase-inc} presents the region of the stable AF phase
in the plane of $t_{pp}$ and $\Delta_{dp}=\epsilon_p-\epsilon_d$. 
For large $\Delta_{dp}=\epsilon_p-\epsilon_d$, we have the region of the AF 
state with an eight-lattice periodicity in accordance with the results of
neutron-scattering measurements\cite{tra96,wak00}.
In the incommensurate antiferromagnetic region, we obtain a finite
SC condensation energy, assuming a spatial oscillation, which is shown in 
Fig. \ref{E-16x4}.
The variational parameters are $g=0.386$, $\tilde{\epsilon}_d=-1.578$,
$\tilde{\epsilon}_p=0$, $\mu=-3.09$, $\Delta_{inc}=0.5$, and
$\Delta=0.02$.

The main results of this study are shown in Fig. \ref{dE-N} where
the size dependence of the SC condensation energy is shown for $x=0.2$, 0.125,
0.08333, and 0.0625.
We set the parameters to be $\epsilon_p-\epsilon_d=2$ and $t_{pp}=0.4$ in
$t_{dp}$ units, which is reasonable from the viewpoint of the density of states
and in the region of eight-lattice periodicity at $x=1/8$.
We have carried out the Monte Carlo calculations up to $16\times 16$ unit cells
(768 atoms in total).  
In the overdoped region in the range of $0.18<x<0.28$,
we have the uniform $d$-wave pairing state as the ground state.
The periodicity of spatial variation judged from the condensation energy  
increases proportionally to $1/x$ as the doping rate $x$
decreases.  In the figure, we have the 12-lattice
periodicity at $x=0.08333$ and the 16-lattice periodicity at $x=0.0625$.
For $x=0.2$, 0.125, and 0.08333, the results strongly suggest a finite condensation
energy in the bulk limit.
We believe that the size dependence of the SC condensation energy in the
incommensurate region is rather weak because the main part of the superfluid
density is in the hole-rich region of the striped structure.
Thus we expect a finite condensation energy even at $x=0.08333$ and 0.0625.
The SC condensation energy obtained on the basis of specific heat
measurements agrees well with the result of variational Monte Carlo 
computations\cite{lor93}.
In general, the Monte Carlo statistical errors are much larger than those
for the single-band Hubbard model.  A large number of Monte Carlo 
steps (more than 5.0$\times 10^7$) is required to obtain convergent expectation
values for each set of parameters.

In Fig. \ref{d-p-phase}  the order parameters $\Delta_{AF}$ and $\Delta_{SC}$ were
evaluated using the formula $E_{cond}=(1/2)N(0)\Delta^2$ where
$N(0)$ is the density of states.  
The SC condensation energy decreases as the doping rate $x$ is decreased
because of the striped structure of the electronic state.
Hence, $\Delta_{SC}$ also decreases.
Here, we have set $N(0)\sim 5/t_{dp}$,
since $N(0)$ is estimated to be $N(0)\sim 2$ to 3 $(eV)^{-1}$ for the 
optimally doped YBa$_2$Cu$_3$O$_{6+x}$ using $N(0)(k_BT_c)^2/2$\cite{and98}.
The phase diagram is consistent with the recently reported phase
diagram for layered cuprates\cite{muk06}.
Although the incommensurate order has never been observed in experiments,
there is a possibility that observed commensurability of magnetic order
may be brought about by the effect of nearby layers, that is, the cancellation of
incommensurability between layers.


We examined the phase diagram of high-temperature superconductors  
with respect to the carrier density,
on the basis of the d-p model.  We carried out variational
Monte Carlo calculations for the 2D d-p model to investigate the ground
state for large $U_d$.  
In the lightly doped region we obtain the coexistent state of 
antiferromagnetism and superconductivity at the doping rate $x=0.125$, 
0.0833 and 0.0625.
As long as we employ the Gutzwiller ansatz, the ground state exhibits
coexistence in the lightly doped region.
In recent experimental works
for layered cuprates, the possibility of the coexistent state of 
antiferromagnetism
and superconductivity has been explored\cite{muk06,cri07}.

  We express our sincere thanks to J. Kondo, S. Koikegami and S. Koike for 
helpful discussions.
This work was supported by a Grant-in Aid for Scientific Research from the
Ministry of Education, Culture, Sports, Science and Technology of Japan.
Parts of the numerical calculations were performed at facilities of the
Supercomputer Center of the Institute for Solid State Physics, University 
of Tokyo, and the Supercomputer Center of High Energy Accelerator Research
Organization (KEK).


\begin{thebibliography}{}

\bibitem{dag94}E. Dagotto: Rev. Mod. Phys. 66 (1994) 763.
\bibitem{ben03}{\em The Physics of Superconductor} Vol.II
edited by K. H. Bennemann and J. B. Ketterson (Springer-Verlag, Berlin,
2003).
\bibitem{and97}P. W. Anderson: {\em The Theory of Superconductivity in
the High-T$_c$ Cuprates} (Princeton University Press, Princeton, 1997).
\bibitem{mor00}T. Moriya and K. Ueda: Adv. Phys. 49 (2000) 555.
\bibitem{hir89}J. E. Hirsch, E. Y. Loh, D. J. Scalapino, and S. Tang:
Phys. Rev. B39 (1989) 243.
\bibitem{sca91}R. T. Scalettar, D. J. Scalapino, R. L. Sugar, and
S. R. White: Phys. Rev. B 44 (1991) 770.
\bibitem{tak97} T. Takimoto and T. Moriya: J. Phys. Soc. Jpn. 66 (1997) 2459.
\bibitem{gue98}M. Guerrero, J. E. Gubernatis, and S. Zhang: Phys. Rev. B 57
(1998) 11980.
\bibitem{kob98}A. Kobayashi, A. Tsuruta, T. Matsuura and Y. Kuroda:
J. Phys. Soc. Jpn. 67 (1998) 2626.
\bibitem{koi00}S. Koikegami and K. Yamada: J. Phys. Soc. Jpn. 69 (2000) 768.
\bibitem{yan01}T. Yanagisawa, S. Koike and K. Yamaji: Phys. Rev. B 64 (2001) 184509.
\bibitem{koi01}S. Koikegami and T. Yanagisawa: J. Phys. Soc. Jpn. 70 (2001) 3499 (2001);
J. Phys. Soc. Jpn. 71 (2002) 671.
\bibitem{yan08}T. Yanagisawa: New J. Physics 10 (2008) 023014.
\bibitem{gro87}C. Gros, R. Joynt, and T. M. Rice: Phys. Rev. B 36 (1987) 381.
\bibitem{yok87}H. Yokoyama and H. Shiba: J. Phys. Soc. Jpn. 56 (1987) 1490.
\bibitem{nak97}T. Nakanishi, K. Yamaji, and T. Yanagisawa: 
J. Phys. Soc. Jpn. 66 (1997) 294.
\bibitem{yam98}K. Yamaji, T. Yanagisawa, T. Nakanishi, and S. Koike:
 Physica C 304 (1998) 225.
\bibitem{tra96}J. Tranquada, J. Axe, D. Ichikawa, N. Nakamura, Y. Uchida,
and B. Nachumi: Phys. Rev. B 54 (1996) 7489.
\bibitem{eme87}V. J. Emery: Phys. Rev. Lett. 58 (1987) 2794.
\bibitem{esk89}H. Eskes, G. A. Sawatzky, and L. F. Feiner: Physica C 160 (1989)
424.
\bibitem{hyb90}M. S. Hybertson, E. B. Stechel, M. Schl\"{u}ter, and
D. R. Jennison: Phys. Rev. B 41 (1990) 11068.
\bibitem{mcm90}A. K. McMahan, J. F. Annett, and R. M. Martin: Phys. Rev.
B 42 (1990) 6268.
\bibitem{him02}A. Himeda, T. Kato, and M. Ogata: Phys. Rev. Lett. 88 (2002) 117001.
\bibitem{miy02}M. Miyazaki, T. Yanagisawa, and K. Yamaji: J. Phys. Chem. Solids 63
(2002) 1403.
\bibitem{gia90}T. Giamarchi and C. Lhuillier: Phys. Rev. B 42 (1990) 10641.
\bibitem{yan02}T. Yanagisawa, S. Koike, and K. Yamaji, J. Phys.: 
Condens. Matter 14 (2002) 21.
\bibitem{yan03}T. Yanagisawa, S. Koike, S. Koikegami, and K. Yamaji: 
Phys. Rev. B 67 (2003) 132408. 
\bibitem{miy04}M. Miyazaki, K. Yamaji, and T. Yanagisawa: 
J. Phys. Soc. Jpn. 73 (2004) 1643.
\bibitem{wak00}S. Wakimoto, R. J. Birgeneau, Y. Endoh, P. M. Gehring, K. Hirota,
M. A. Kastner, S. H. Lee, Y. S. Lee, G. Shirane, S. Ueki, and K. Yamada: 
Phys. Rev. B 61 (2000) 3699.
\bibitem{lor93}J. W. Loram, K. A. Mirza, J. R. Cooper, and W. Y. Kiang: 
Phys. Rev. Lett. 71 (1993) 1740.
\bibitem{and98}P. W. Anderson: Science 279 (1998) 1196.
\bibitem{muk06}H. Mukuda, M. Abe, Y. Araki, Y. Kitaoka, Y. Tokiwa, T. Watanabe,
A.Iyo, H. Kito, Y. Tanaka: Phys. Rev. Lett. 96 (2006) 087001.
\bibitem{cri07}A. Crisan, Y. Tanaka, A. Iyo, D. D. Shivagan, P. M. Shirage,
K. Tokiwa, T. Watanabe, L. Cosereanu, T. W. Button, and J. S. Abell: 
Phys. Rev. B 76 (2007) 212508.
 
\end{thebibliography}
\end{document}